\documentclass[floatfix,%
 reprint,
 amsmath,amssymb,
 aps,
]{revtex4-2}
\usepackage{graphicx}
\usepackage{dcolumn}
\usepackage{bm}
\usepackage[colorlinks=true, linkcolor=black, citecolor=blue, urlcolor=black]{hyperref}
\usepackage{xcolor}
\usepackage{amsmath}
\usepackage{amssymb}

\def\md{\mbox{d}}
\def\Aav{\langle A \rangle}

\def\NA{\langle Z_A \rangle}

\begin{document}

\title{Sequential binding-unbinding based specific interactions influence exchange dynamics and size distribution of protein condensates}
\thanks{Supplementary information (SI) available}%

\author{Bhanjan Debnath}
\author{Parag Katira}%
 \email{Corresponding author: bdebnath@sdsu.edu; pkatira@sdsu.edu}
\affiliation{%
 Department of Mechanical Engineering, San Diego State  University, CA, 92182, USA
}

\begin{abstract}
The interaction lifetimes between condensate-forming biomolecules can dictate both the specificity of the condensate-forming species as well as the fluidity and exchange dynamics of these condensates. Using a heuristic modeling approach, we show that single-step vs. sequential, multistep binding-unbinding interactions between proteins can lead to similar average interaction lifetimes, but with either exponential or truncated power-law-like lifetime distributions, respectively. Combining this model with Brownian dynamics simulations, we find that the differences in these lifetime distributions influence the features of condensates, such as their fluidic nature, aging, and size distribution. Overall, we highlight a key, but often ignored aspect of biomolecular condensate forming processes--the lifetime distributions of individual interactions may singularly control  condensate dynamics. 
\end{abstract}

\maketitle

{\it Introduction---}The formation of organelles and biomolecular condensates has been observed in numerous biological processes inside cells \citep{hyman2014liquid, shin2017liquid,banani2017biomolecular}. Such membraneless compartments, for example, P-granules and stress granules, exhibit droplet-like features \citep{brangwynne2009germline,patel2015liquid}. Their common features include internal rearrangement, material exchange, coarsening, dissolution, etc. \citep{banani2016compositional,caragine2019nucleolar,lee2023size, siggia1979late,tanaka1995new, tanaka2000viscoelastic,shimizu2015novel}. These features are essential for crucial biological functions, such as signaling and maintaining a biochemical environment within organelles by sequestering unwanted molecules. But, not all condensates need to remain in a liquid-like state. The transition from a liquid to a gel-like state has been observed as the condensate ages or matures through solvent expulsion \citep{jawerth2020protein,vidal2022liquid,poudyal2023intermolecular,emmanouilidis2024solid}. These transitions are linked to cataract, pathological aggregation, and neurodegenerative diseases \citep{benedek1997cataract,patel2015liquid,mehra2019alpha,ray2020alpha}. The growing number of biological processes in which the involvement of biomolecular condensates has been observed, with limited understanding of their roles, fuels the intrigue and interest in condensate dynamics \citep{shin2017liquid,banani2017biomolecular}. 

In the context of protein-protein interactions, it is well understood how strong and weak interactions among proteins containing intrinsically disordered regions are responsible for the formation of condensates \citep{brangwynne2015polymer,choi2020physical,zhang2021decoding,poudyal2023intermolecular,hess2025structured}. In the context of the features of condensates, many attempts have been made to identify the factors that influence the physical properties of liquid-like condensates, such as their coarsening dynamics, fluidic nature, material exchange dynamics, size distribution,  the transition from a liquid-like to a gel-like state, and aging \citep{lin2022modeling,garaizar2022aging,takaki2023theory,lee2023size,zhang2024exchange,biswas2024molecular}. However, factors that drive the formation of protein-specific condensates inside a crowded, multicomponent cellular environment remain to be fully elucidated. In addition, very few studies have attempted to establish the link between protein-protein interactions and condensate features \citep{sundaravadivelu2024sequence,hess2025structured}, and how changes in these interactions alter condensate dynamics.  

One plausible factor contributing to the features of condensates is `the lifetime of protein interactions', which, although seemingly intuitive, is less questioned and can play a crucial role in regulating condensate dynamics. We define the lifetime of interactions as the duration for which any two protein molecules in close proximity remain to be continuously bound. If these durations are extremely long, this may indicate that the condensates behave as solids, whereas very short-lived interactions may not allow the formation of stable condensates, resulting in rapid dissolution. Intuitively, one would expect a trade-off between longer interaction lifetimes, which lead to interaction specificity between condensate-forming molecules, and the fluidity of the condensates. Protein binding sites, binding strengths, thermal fluctuations, and the strengths of external forces impact the relationship between interaction lifetimes and condensate dynamics. Consequently, two key questions arise: 1) how the mechanisms of binding and unbinding between interacting proteins give rise to distinct interaction lifetimes, and 2) how the nature of lifetime distributions affect the dynamics of condensate formation while maintaining the specificity (for example, the same mean lifetime) between the interacting proteins or biomolecules.

Over the decades, various mechanisms for binding-unbinding between protein-protein and protein-surface have been proposed \citep{katira2009random,changeux2011conformational,seo2014protein,stank2016protein}. Binding-unbinding is generally considered a single-step, first-order reversible reaction process, leading to exponential lifetime distributions. However, the former may not be true if binding-unbinding do not follow a single-step pathway \citep{biancaniello2005colloidal,rogers2013kinetics,penna2014molecular,armstrong2020power,garcia2021power}, whence the lifetime distributions are non-exponential. Therefore, in the context of condensate formation, the kinetics of binding and unbinding may be of significant relevance. These kinetics can drive the specificity in protein interactions \citep{chen2019programmable}, thus contributing to the formation of protein-specific condensates. Here, we explore whether they can also influence the features of the condensates.

In this work, we model selective interactions as either a single-step or a sequential, multi-step binding-unbinding process. We describe a mechanism that elucidates the dynamics of binding and unbinding, providing insight into interactions with high specificity. We use a minimal hybrid framework that integrates these interactions with Brownian dynamics to simulate the motion of individual protein molecules and their clusters. Our model captures key features of condensate and demonstrates how specific binding-unbinding interaction pathways influence condensate features, highlighting the link between the kinetics of protein interactions and condensate dynamics.

\begin{figure}
\centering
\includegraphics[width=1\linewidth]{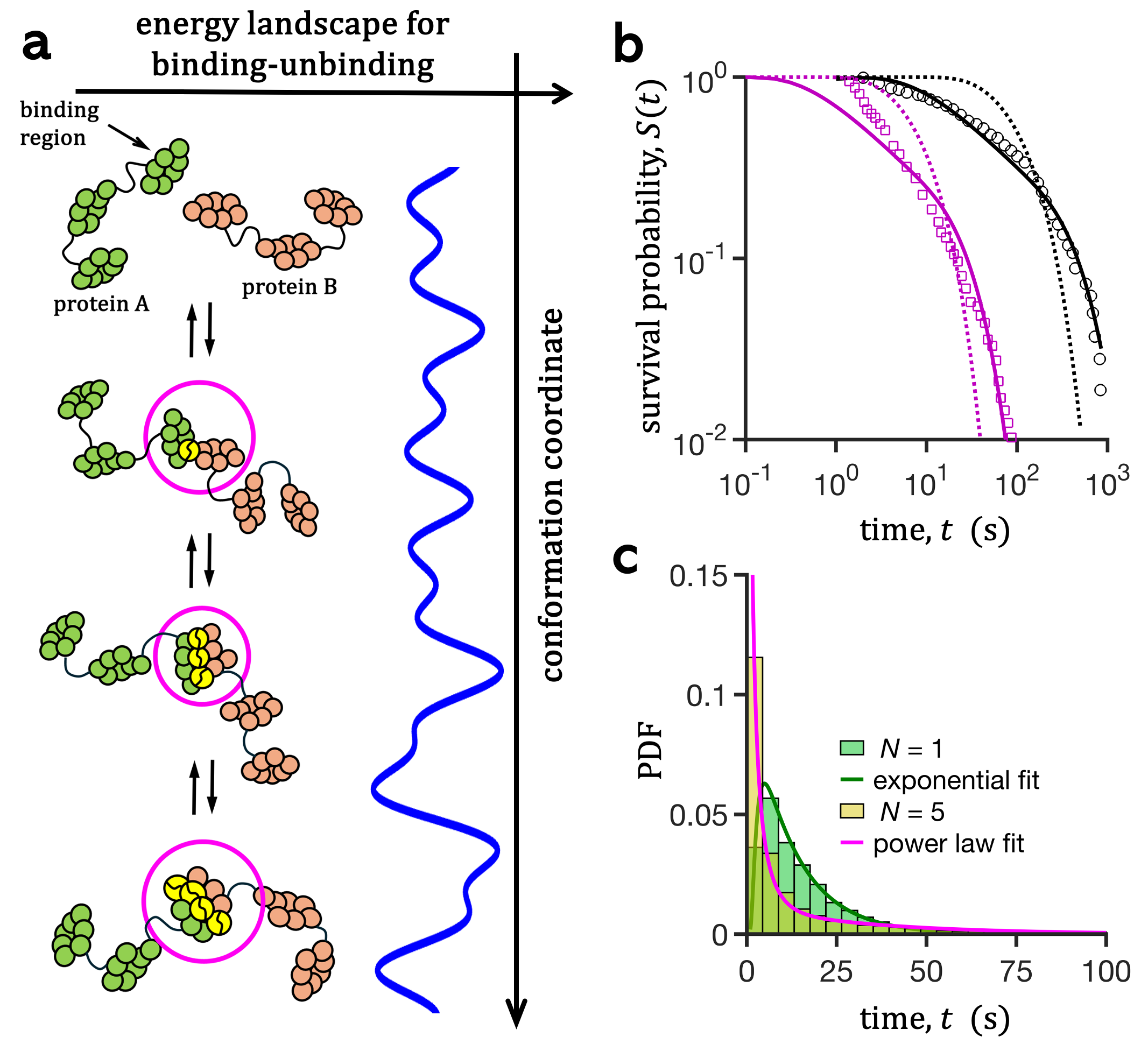}
\caption{(a) Schematic of sequential binding-unbinding interaction. Binding and unbinding through conformational changes occur in a sequential manner on the conformation coordinate. (b) Survival probability $S(t)$ with time $t$, and comparison of model predictions with experiments of Armstrong \textit{et al.} ($\circ$) \citep{armstrong2020power} and Garcia \textit{et al.} ($\square$) \citep{garcia2021intrinsically}. The black and pink curves are the model predictions and best fits to the data $\circ$ and $\square$, respectively. The black solid and dotted curves correspond to \{$N = 5$, $\tilde{\mathcal{C}} = 0.03$\} and \{$N = 1$, $\tilde{\mathcal{C}} = 0.0038$\}, respectively, with the same mean lifetime $\tau_m = 130$ s. The pink solid and dotted curves correspond to \{$N = 5$, $\tilde{\mathcal{C}} = 0.5$\} and \{$N = 1$, $\tilde{\mathcal{C}} = 0.055$\}, respectively, with  $\tau_m = 9$ s.  Here $\tilde{\mathcal{C}} = \mathcal{C}/\alpha$ and $\alpha$ represents one square unit of conformational changes per  second.  (c) The lifetime distributions (PDF = $- \, \md S/\md t$) for $N = 5$ and $N = 1$; the choices of $\tilde{\mathcal{C}}$ are such that $\tau_m$ of both distributions is the same. In (c), for $\tau_m = 15$ s, $\tilde{\mathcal{C}} = 0.3$ and $\tilde{\mathcal{C}} = 0.034$ for  $N = 5$ and $N = 1$, respectively.} 
\label{fig:schematic}
\end{figure}

{\it Model---}Except for long-range and electrostatic attractions, the high specificity in protein-protein interactions can arise because of their backbone shape and `extensive and
modular side-chain hydrogen-bond networks' \citep{chen2019programmable,baker2019has}. These networks become gradually accessible for binding and unbinding through conformational changes \citep{lu2005probing,guo2016protein}. These selective interactions are often coined as `interaction specificity', especially when the protein molecules are in a heterogeneous and multicomponent chemical environment \citep{hess2025structured}. Modeling such intricate interactions when two protein molecules are in close proximity is challenging, prompting us to adopt a heuristic approach.

We assume that binding and unbinding through conformational changes occur in a sequential manner (see Fig.~\ref{fig:schematic}a). 
For simplicity, we further assume that the probability in a future conformational bound state $n_{(t+\Delta t)}$  depends on the current bound state $n_{t}$, i.e.,  $\mathcal{P}(n_{(t+\Delta t)}|n_{t}, n_{(t-\Delta t)}, \dots, n_{t=0}) = \mathcal{P}(n_{(t+\Delta t)}|n_{t})$. If there is a bias that favors certain bound states, then the system evolves as  
\begin{align}
    \frac{\partial \mathcal{P}(n,t)}{ \partial t} 
    =  \frac{\partial}{ \partial n}\Big[\mathcal{C}(n,t) \, \frac{\partial \mathcal{P}(n,t)}{\partial n} - \mathcal{U}(n,t) \, \mathcal{P}(n,t)\Big],
    \label{eq:master}
\end{align}
where $\mathcal{U}(n,t)$ is the drift, and $\mathcal{C}(n,t)$ is the diffusivity associated with conformational changes. Both can be related to the difference in the height of the energy barrier and the depth of the energy well for transitioning between different conformational bound states. Although more work is required to generate energy landscapes for specific interactions \citep{brooks2001taking,wales2006potential,joseph2017exploring}, and to develop relations for $\mathcal{U}(n,t)$ and $\mathcal{C}(n,t)$ in connection with energy landscapes, this is beyond the scope of the present work. Here, we now focus on simplified cases. The survival probability, $S(t) = \int \mathcal{P}(n,t) \, \md n$, represents the likelihood that the bound protein molecules remain attached after a period of time. Therefore, $(-\md  S/\md t)$ corresponds directly to the lifetime distribution of these specific interactions. We note in passing that we prefer to decouple the effects of other long-range and electrostatic interactions, and hence, they are ignored here.

As Eq.~\ref{eq:master} resembles the standard advection-diffusion equation, an exact solution of Eq.~\ref{eq:master} can be obtained for a special case without drift and no variation of $\mathcal{C}$ with $n$ and $t$, corresponding to a uniform energy landscape. For simplicity, we adopt this case to test the potential and predictability of the model. For this special case and a total number of probable conformational changes $N$, $S(t)$ is obtained as (see the supplementary information (\textit{SI}) section SI1)
\begin{align}
    S(t) = \frac{4}{\pi} \, \sum_{\ell = 1}^{\infty} \, \Bigg\{ \frac{1}{(2 \ell - 1)}  \, \, {\rm sin}\Bigg[\frac{(2 \ell - 1) \pi}{2 N}\Bigg] \, \, \nonumber \\
    \times \,  {\rm exp}\Bigg[- \, \mathcal{C} \, \Big(\frac{(2 \ell - 1) \pi}{2 N}\Big)^2 \, t\Bigg] \Bigg\}.
    \label{eq:master_2}
\end{align}

This model captures the truncated power-law trend observed in experiments reasonably well when binding-unbinding is a sequential multi-step process, i.e., $N > 1$ (Fig.~\ref{fig:schematic}b). The experimental data represent protein (fibrinogen) -- surface (glass) interactions \citep{armstrong2020power} and protein-protein (transcription factors) interactions involving intrinsically disordered regions \citep{garcia2021intrinsically}. However, when the binding-unbinding is considered as a single-step process ($N = 1$), the model instead predicts an exponential trend, which does not capture the data satisfactorily (Fig.~\ref{fig:schematic}b). 
Truncated power-law lifetime distributions have been reported earlier in a variety of biological and bio-derived systems, such as protein-DNA (or RNA) strand interactions, protein-protein interactions, protein-surface interactions and between biomolecule-coated microspheres \citep{biancaniello2005colloidal,rogers2013kinetics,penna2014molecular,mazzocca2021transcription,elkins2021direct, garcia2021power,lyu2022protein}. This suggests that sequential multi-step binding-unbinding interactions may be more prevalent across biological contexts than previously appreciated. Nevertheless, the former does not imply that the power-law behavior is universal for all types of proteins \citep{kulin2002real}. 
Therefore, a thorough validation and calibration of $N$ and $\mathcal{C}$ for different protein types would require studying their interactions through controlled experiments \citep{gebhardt2013single, banerjee2018single, bespalova2019, bespalova2022,walker2024sensing} or computationally expensive all-atom MD simulations \citep{henry2013comparing,dommer2025all}, which merits further attention. 

Interestingly, despite the same mean lifetime $\tau_m$, the two types of specific binding-unbinding interactions ($N = 1$ and $N > 1$) result in distinct lifetime distributions (Fig.~\ref{fig:schematic}c). In Fig.~\ref{fig:schematic}c, both distributions indicate that short and long lifetimes are probable. Though the probability of short lifetimes is higher for $N > 1$ than that of $N = 1$, the fat-tail for $N > 1$ serves to complement the mean lifetime $\tau_m$ (see \textit{SI} section SI1 and Fig. S1). 

We now integrate this model with Brownian dynamics simulations in a hybrid manner to examine the influence of such specific binding-unbinding interactions on the features of condensates. In the simulations, both individual molecules and clusters are allowed to diffuse. Each molecule has a certain number of binding regions, represented by $\mathcal{R}_B$. When two molecules (whether part of clusters or individuals) are in close proximity, they bind with a sticking probability $p_s$.  The molecules remain bound for a lifetime that is randomly drawn from the PDFs shown in Fig.~\ref{fig:schematic}c, followed by unbinding (see the simulation details in \textit{SI} section SI2.). 
The long-range hydrodynamic interactions are ignored here.

\begin{figure}
\centering
\includegraphics[width=1\linewidth]{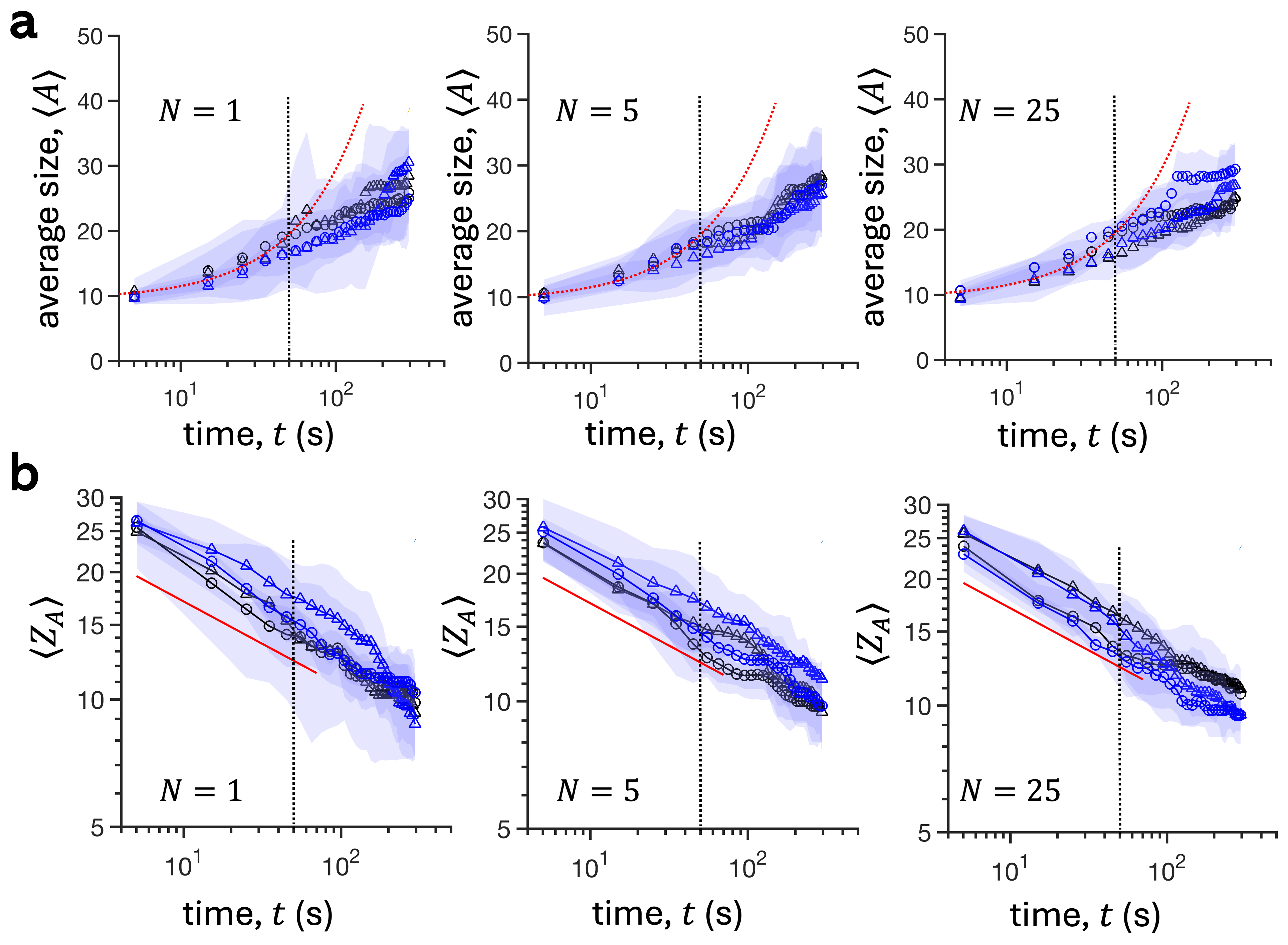}
\caption{Variation of the average size of clusters $\Aav$ (a) and the total number of clusters $\NA$ (b) with time $t$ for different values of the total number of probable conformational changes $N$. Here, the black and blue symbols are for the number of binding regions $\mathcal{R}_b = 3$ and $\mathcal{R}_b = 4$, respectively, and  $\circ$ and $\triangle$ represent the mean lifetime of the lifetime distributions $\tau_m = 15$ s and $\tau_m = 130$ s, respectively. Panels in (a) and (b) correspond to 
the area fraction $a_f = 0.1$. The red dotted curves in panel (a) fit the data points by $y = (mx + c)$ on the left of the black dotted vertical lines. The red solid lines in panel (b) represent the slope $-0.2$. The shaded regions represent the 95\% confidence limits from four independent simulations.} 
\label{fig:size_time}
\end{figure}

{\it Results---}We define the average size of clusters as $\Aav \, = \, (\Sigma_{A\geq2} \, z_A \, A^2)/(\Sigma_{A\geq2} \,  z_A \, A)$ and total number of clusters as $\NA = \Sigma_{A\geq2} \, z_A$  \citep{vicsek1984dynamic,family1985cluster}, where $z_A$ is the number of clusters of size $A$. Here, $A$ is the two-dimensional volume of a cluster scaled by the nominal two-dimensional volume $a_p$. Without loss of generality, $a_p$ is set to 1. The growth of $\Aav$ over time shows diffusion-limited aggregation in the early stage of phase separation (Fig.~\ref{fig:size_time}a and \textit{SI} Fig. S3a). In this period, the variation of $\NA$ with time shows a linear trend on the log-log scale (Fig.~\ref{fig:size_time}b and \textit{SI} Fig. S3b). It is evident that molecular collisions via diffusion is one of the prerequisites for cluster formation, growth, and aggregation. We could have expected reaction-limited growth if the binding kinetics (incorporated as the sticking probability $p_s$ in the model) had been much slower than diffusion, implying the requirement of numerous collisional contacts before a successful binding event \citep{family1985cluster,lin1990universal}. Family \textit{et al.} \citep{family1985cluster} showed that cluster aggregation is chemically controlled and reaction-limited if $p_s \ll 0.1$. However, the present study employs $p_s = 0.1$ and therefore we can expect diffusion-limited growth, consistent with the findings of Family \textit{et al.} \citep{family1985cluster}.  

In pure growth, if the coarsening of molecules and clusters follows the Smoluchowski coagulation dynamics, then $\Aav \propto t$ in the absence of hydrodynamic interactions \citep{chandrasekhar1943stochastic,siggia1979late,kumaran1998droplet,kumaran1998effect,tanaka2000viscoelastic,chen2024fractal}. This scaling is reflected in Fig.~\ref{fig:size_time}a and \textit{SI} Fig. S3a, irrespective of the ranges of parameters. However, the coarsening dynamics deviates from this scaling in the long-time limit. In the variation of $\NA$ with time, there are minute oscillations of small amplitudes and large wavelengths in the long-time limit (Fig.~\ref{fig:size_time}b and \textit{SI} Fig. S3b). This is an indication of aggregation-fragmentation processes \citep{brilliantov2018steady}. The fragmentation process is driven by the unbinding events. In that case, the growth dynamics does not necessarily need to follow the same scaling as in diffusion-limited growth in the long-time limit. The findings of simultaneous growth, aggregation, and fragmentation in Fig.~\ref{fig:size_time} collectively imply that clusters are dynamic, and they can participate in neighbor exchange events.

\begin{figure}
\centering
\includegraphics[width=0.85\linewidth]{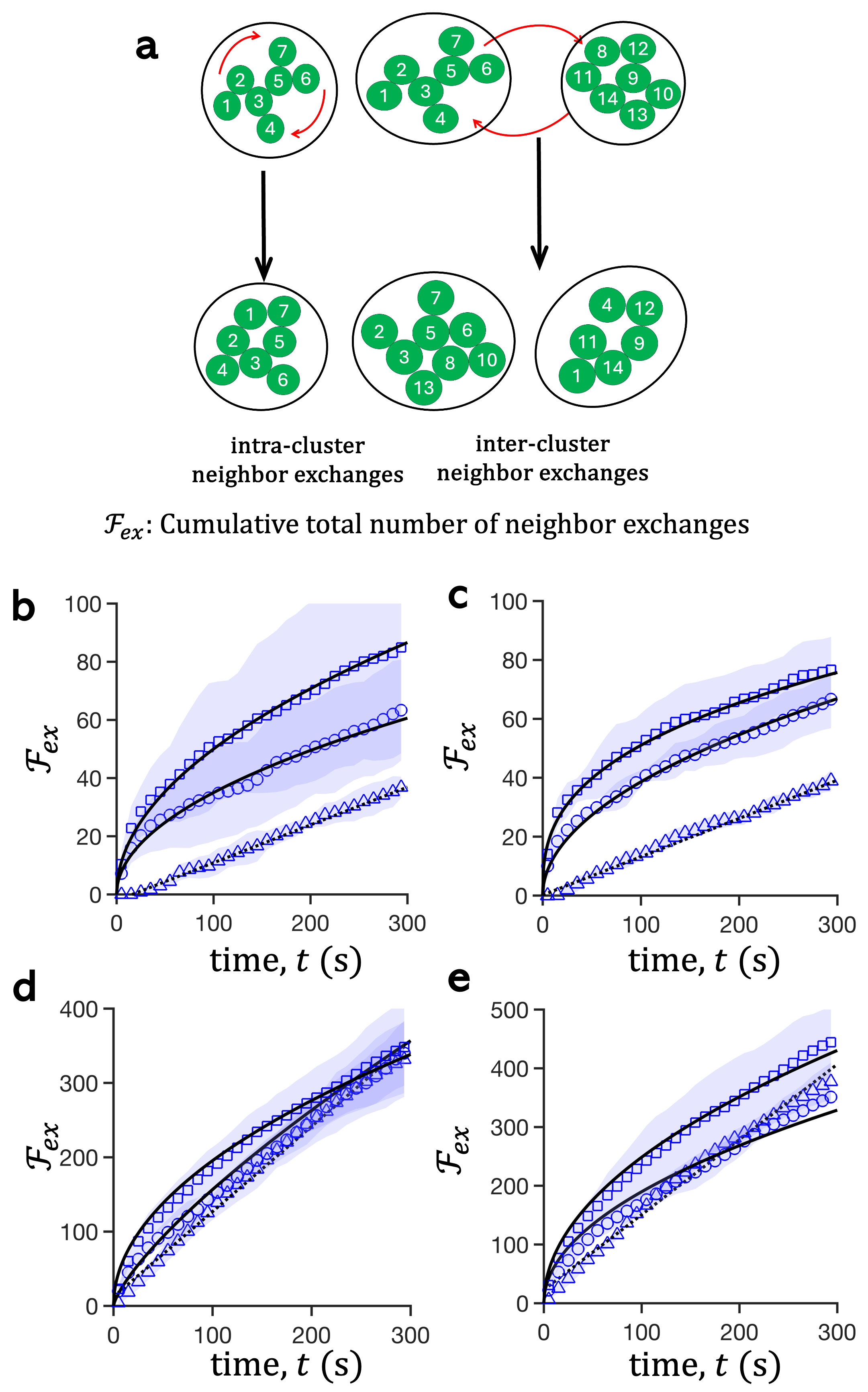}
\caption{(a) Schematic of different types of neighbor exchanges. Variation of the cumulative total number of exchanges $\mathcal{F}_{ex}$ with time $t$. Parameters: (b) area fraction $a
_f = 0.05$, the mean lifetime of distribution $\tau_m = 130$ s; (c) $a
_f = 0.1$, $\tau_m = 130$ s; (d) $a_f = 0.05$, $\tau_m = 15$ s; (e) $a_f = 0.1$, $\tau_m = 15$ s. In (b)--(e), the data correspond to $\mathcal{R}_b = 3$. Results for $\mathcal{R}_b = 4$ and comparison with $\mathcal{R}_b = 3$ has been shown in \textit{SI} Fig. S4.   Here, $\triangle$, $\circ$, and $\square$ are for the total number of probable conformational changes $N = 1$, $N = 5$, and $N = 25$, respectively. The black solid curves are the fits for $y = m x^{1/2}$ and black dotted lines for $y = m x$. The shaded regions represent the 95\% confidence limits from four independent simulations.} 
\label{fig:exchanges_time}
\end{figure}

A schematic illustrating the different modes of neighbor exchange is shown in Fig.~\ref{fig:exchanges_time}a. Despite the same mean lifetime $\tau_m$, the cumulative total number of neighbor exchange events $\mathcal{F}_{ex}$ (see \textit{SI} section SI3 for how to obtain $\mathcal{F}_{ex}$) is strongly affected by the changes in the binding-unbinding pathways due to $N$, and the number binding regions per molecule $\mathcal{R}_b$ (Fig.~\ref{fig:exchanges_time}b,c and \textit{SI} Fig. S4b,c). During the early stage of phase separation, $\mathcal{F}_{ex}$ is higher for $N > 1$ than for $N = 1$. This occurrence is expected as the probability of very short lifetimes is higher and the probability of intermediate lifetimes is lower for $N > 1$, compared to that for $N = 1$ (see Fig.~\ref{fig:schematic}c). In Fig.~\ref{fig:exchanges_time}b,c and \textit{SI} Fig. S4b,c, as $\tau_m = 130$ s is comparable to the total duration of the simulations, which is 300 s, it might initially appear that $\mathcal{F}_{ex}$ would continue to increase with time. However, the scalings $ \mathcal{F}_{ex} \propto t^{1/2}$ for $N > 1$ and $ \mathcal{F}_{ex} \propto t$ for $N = 1$ imply that $\mathcal{F}_{ex}$ would tend to saturate for $N > 1$, but $\mathcal{F}_{ex}$ would continue to increase for $N = 1$ over time $t \gg \tau_m$. Therefore, a crossover between $N = 1$ and $N > 1$ would be expected in time $t \gg \tau_m$. This is indeed the case when $\tau_m$ is reduced to 15 s, which is much smaller than the total duration of the simulations; the crossover is noticeable in the long-time limit (Fig.~\ref{fig:exchanges_time}d,e and \textit{SI} Fig. S4d,e). 

The findings in Fig.~\ref{fig:exchanges_time} reveal that the system with a sequential, multi-step binding-unbinding process ($N > 1$) shows a higher rate of exchange events during the early stages of phase separation. However, the former system also undergoes aging through a gradual decrease in the frequency of exchange events in the long-time limit.  Interestingly, for the same $\tau_m$, one mechanism ($N > 1$) leads to aging, while another ($N = 1$) does not. These findings indicate a connection between aging and changes in protein-protein interactions, which has previously been speculated as a time-dependent evolution in strength and specificity through minor structural and conformational changes in protein interactions
\citep{patel2015liquid, hess2025structured}.  

Note that the prolonged lifetimes can arise when $N > 1$ (see Fig.~\ref{fig:schematic}c and \textit{SI} Fig. S1). 
This scenario can be visualized as two protein molecules sequentially binding to one another, becoming trapped in a conformational state as
$n \rightarrow  N$ (see Fig.~\ref{fig:schematic}a). The protein molecules then would probably take longer time to return to the unbound state before participating in exchange events. This trapping mechanism is analogous to the continuous-time random-walk trap model described by Bouchaud \citep{bouchaud1992weak,monthus1996models}. Recently, similar trap models have been developed to explain the rheology of aging condensates \citep{lin2022modeling,takaki2023theory}. For $N = 1$, the time intervals between successive neighbor exchange events would be exponentially distributed, as in a Poisson process. With a mean waiting time $\tau_m$, the number of jumps within time $t$ would be scaled as $\sim t/\tau_m$, and thus $ \mathcal{F}_{ex} \propto t$ for  $N = 1$. But, if there are traps in the jump process, leading to prolonged waiting times, the distribution of jump times would be a heavy-tailed power-law, similar to the continuous-time random-walk trap model \citep{monthus1996models}. Then the number of jumps within time $t$ would be scaled as $\sim t^{1/2}$ for L\'evy distribution which is an outcome of a limiting case $N \rightarrow \infty$. Therefore, we can expect the scaling $ \mathcal{F}_{ex} \propto t^{1/2}$ for $N > 1$. 

\begin{figure}
\centering
\includegraphics[width=0.9\linewidth]{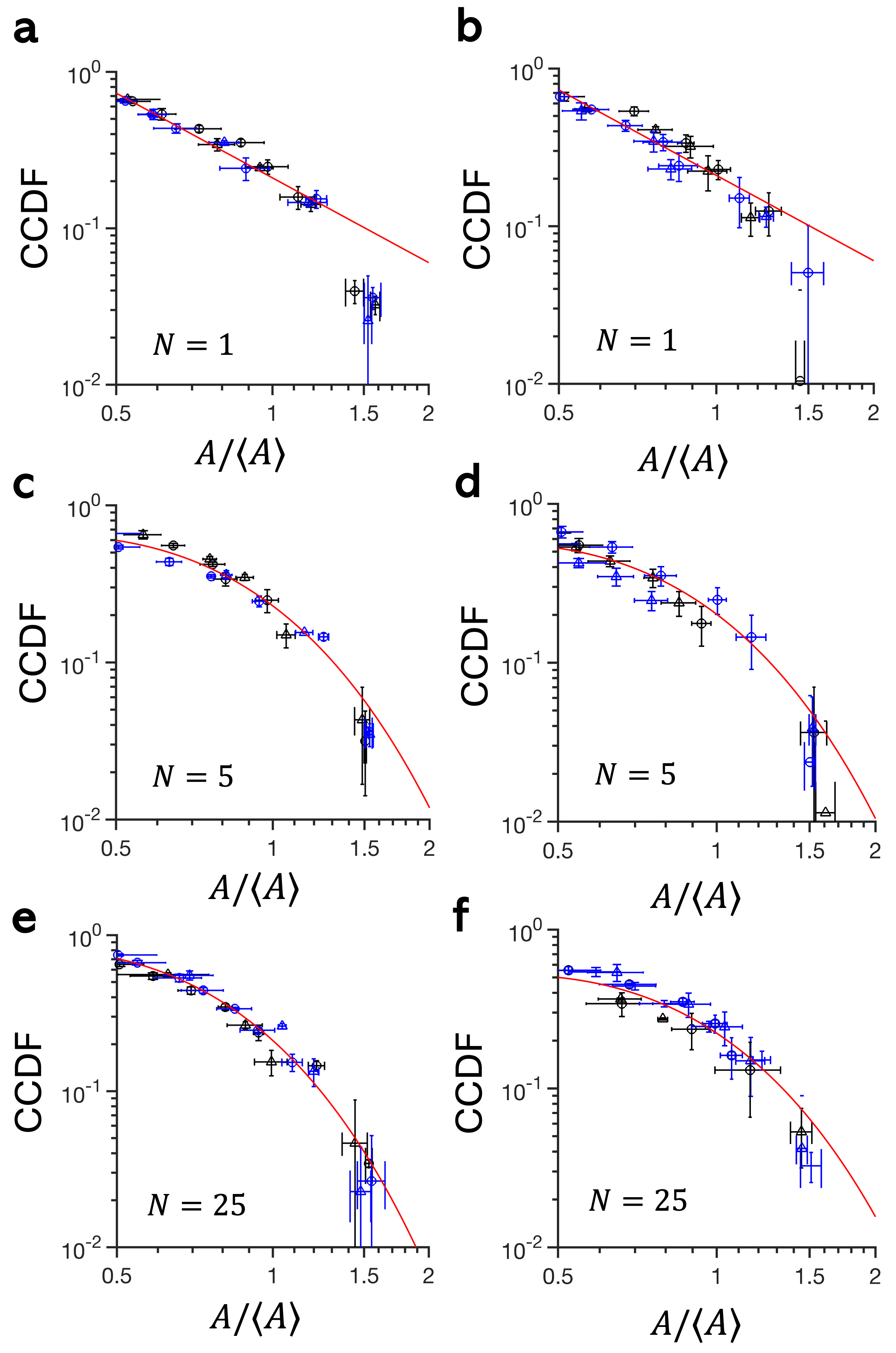}
\caption{After $t =  300$ s of simulation, the complementary cumulative distribution functions (CCDF) of the scaled size $A/\Aav$ of the clusters are shown in log-log scale:  (a), (c), and (e) are for the area fraction $a_f = 0.05$, and (b), (d), and (f) are for $a_f = 0.1$. The black and blue symbols are for the number of binding regions $\mathcal{R}_b = 3$ and $\mathcal{R}_b = 4$, respectively, and  $\circ$ and $\triangle$ represent the mean lifetime of the lifetime distribution $\tau_m = 15$ s and $\tau_m = 130$ s, respectively. The red lines in (a) and (b) are fit for $y = m x^\alpha$, and the red curves in (c)--(f) for $y = a \, {\rm exp}(-bx) \, x^\beta$. The error bars represent the 95\% confidence limits from four independent simulations.} 
\label{fig:size_dstr}
\end{figure}

Plotting the complementary cumulative distribution function (CCDF), where ${\rm CCDF} = 1 - \int_0^A \, f(A^\prime,t)\, \md A^\prime$ and $f(A^\prime,t)$ is the probability density of size $A^\prime$ of clusters, with the scaled size of clusters $A/\Aav$ on the log-log scale, two distinct trends emerge for $N = 1$ and $N > 1$ (Fig.~\ref{fig:size_dstr}). The change in curvature is clearly observed in Fig.~\ref{fig:size_dstr}, when the binding-unbinding pathway changes from $N = 1$ to $N > 1$. Different size distributions have been previously reported in experiments \citep{caragine2019nucleolar, lee2023size}. Nucleation and preferential merging were shown to drive the differences in size distributions \citep{lee2023size}. However,  the role of protein-protein interactions has not been considered explicitly. Recall that we use the sticking probability $p_s$ to account for the same nucleation and merging mechanism of molecules and clusters. Our hybrid model allows us to isolate all other factors \citep{kumaran1998droplet,kumaran1998effect,tanaka2000viscoelastic,shimizu2015novel}, and focus specifically on the link between size distributions and protein-specific interaction mechanisms.
Our findings reveal a link between changes in the size distribution and alteration of protein-protein interactions through structural changes \citep{linsenmeier2022dynamic,hess2025structured}. Connecting protein structural changes to condensate dynamics can provide critical insight into biological processes linked to disease progression or recovery \citep{patel2015liquid, molliex2015phase, shin2017liquid, buchwalter2017nucleolar}.

{\it Discussions---}The specificity in protein interactions required for the formation of protein-specific condensates in a crowded, heterogeneous cellular environment may at times seem at odds with the fluidic nature and dynamic features of the condensates. One explanation of fluidity and rapid exchange of molecules between the condensed (dense) and bulk (dilute) phases involves a tight control of the condensing protein concentration just above the saturation concentration ($C_{\text{sat}}$). Another explanation relies on the presence of biochemical or biomechanical activity within the condensates to supply the energy needed to rapidly breakdown strong interactions. Intuitively, the activity-driven fluidization of condensates appears to be a more likely explanation than a strict concentration control of interacting proteins within biological systems.

In this work, we propose an alternative protein interaction mechanism to explain key features of biological condensates such as fluidity and aging. The explanation focuses on specific binding-unbinding interactions that lead to distinct interaction lifetime distributions. When binding-unbinding follows a single-step process, a common assumption for biochemical reactions, the interaction lifetimes are exponentially distributed. On the other hand, if the binding-unbinding occurs via a sequential, multi-step process due to structural changes of the interacting proteins, the interaction lifetime follows a truncated power-law distribution. Such lifetime distributions have been observed in experiments. However, the origin of these distributions has been attributed to the existence of multiple interacting populations, each with its own unique exponential distribution. Recently, it has been shown by us and others that such lifetime distributions can arise within single populations of interacting proteins as well, when the binding-unbinding process follows a multistep, sequential process. Such non-exponential interaction lifetime distributions are a function of the number of steps within the binding-unbinding process, as well as the mobility of the protein across these quasistable bound conformations. This binding-unbinding mechanism explains the observed biological richness of condensates, for example, their fluidic nature, aging, and size distributions. These results underscore the need to dig deeper into factors such as `high interaction specificity and low conformational heterogeneity' as described by Hess \& Joseph \cite{hess2025structured}, and how together they influence interaction lifetimes and the biophysics of condensates. The proposed interaction mechanism and consequent non-exponential lifetime distributions might also influence other processes as such as enzyme cascades \citep{zhang2017toward,zhao2018substrate} and molecular communication networks \citep{nakano2012molecular}. 

It is important to mention that the present modeling framework needs more refinement to incorporate electrostatic and other long-range interactions, including hydrodynamic interactions. Other important factors that require consideration are reaction-dependent non-equilibrium active processes and diffusiophoretic activity, which can modify the diffusivity of the Brownian droplets, the sticking probability, and the coarsening dynamics. These are the directions for future work.  

\begin{acknowledgments}
The authors thank Henry Hess for valuable discussions. The authors acknowledge funding support from US Army Research Office grants W911NF-23-1-0329, W911NF-22-1-0047, and the San Diego State University Research Foundation.
\end{acknowledgments}

\bibliography{References}
\end{document}


\maketitle

\section*{Supplementary Information (SI)}
\beginsupplement

\subsection{Lifetime distributions from sequential binding-unbinding based model}

Consider the case with a uniform energy landscape for sequential binding and unbinding. Therefore, neglecting the drift and treating $\mathcal{C}$ as a constant, the system is represented by
\begin{equation}
    \frac{\partial \mathcal{P}(n,t)}{ \partial t} 
    =  \mathcal{C} \,\frac{\partial}{ \partial n}\Big[ \frac{\partial \mathcal{P}(n,t)}{\partial n}\Big].
    \label{eq:master_sp}
\end{equation}
We assume that $n = 0$ state is an unbound state. While binding and unbinding, the first conformation bound state is represented by $n = 1$. A total number of probable conformation changes that can occur during sequential binding and unbinding is $n = N$. To solve \ref{eq:master_sp}, we use $n = 0$ and $n = N$ as absorbing and reflective boundaries, respectively. Hence, the boundary conditions used to solve Eq. \ref{eq:master_sp} are
\begin{align}
  \mathcal{P}(n = 0, t) = 0 \\
  \frac{\partial \mathcal{P}}{\partial n} (n = N,t) = 0
    \label{eq:BCs}
\end{align}
We use the following as an initial condition as $n = 1$ would be the first bound state
\begin{align}
  \mathcal{P}(n, t = 0) = \delta(n - 1) \nonumber \\
  \delta(n - 1) = 0 \, \, {\rm for \, \,  all}  \, \, n \, \, {\rm except} \, \, n = 1
    \label{eq:IC}
\end{align}
which satisfies $\int \mathcal{P} \, \md n = \int \delta (n - 1) \, \md n = 1$. Although conformational bound states are represented by discrete whole numbers $\mathcal{W} = \{0, 1, 2, 3, 4, 5, \dots\}$, we solve the above system by treating it as a continuum. An exact solution is possible for the above system \citep{polyanin2001handbook,zaitsev2002handbook},    
\begin{align}
\mathcal{P}(n,t) = \frac{2}{N} \, \sum_{\ell = 1}^{\infty} \, \Bigg\{  {\rm sin}\Bigg[\frac{(2 \ell - 1) \pi}{2 N}\Bigg] \, \,{\rm sin}\Bigg[\frac{(2 \ell - 1) \pi}{2 N} \, n\Bigg] \, \,  \nonumber \\
    \times \,  {\rm exp}\Bigg[- \, \mathcal{C} \, \Big(\frac{(2 \ell - 1) \pi}{2 N}\Big)^2 \, t\Bigg] \Bigg\} 
\end{align}
Next, the survival probability, which is defined as  $S(t) = \int_{n = 0}^{n = N} \, \mathcal{P}(n,t) \, \md n$, is 
\begin{align}
    S(t) = \frac{4}{\pi} \, \sum_{\ell = 1}^{\infty} \, \Bigg\{ \frac{1}{(2 \ell - 1)}  \, \, {\rm sin}\Bigg[\frac{(2 \ell - 1) \pi}{2 N}\Bigg] \, \, \nonumber \\
    \times \,  {\rm exp}\Bigg[- \, \mathcal{C} \, \Big(\frac{(2 \ell - 1) \pi}{2 N}\Big)^2 \, t\Bigg] \Bigg\}.
    \label{eq:sur}
\end{align}
The probability distribution of lifetime can be obtained as PDF  $ = \mathcal{F}(t_i) = -\md S(t)/ \md t $. We obtain the lifetimes of the interaction $t_i$ by taking the inverse, i.e., $t_i = \mathcal{F}^{-1}({\rm PDF})$, and the mean of $t_i$ is represented by the mean lifetime $t_m$.

\begin{figure}
\centering
\includegraphics[width=1\linewidth]{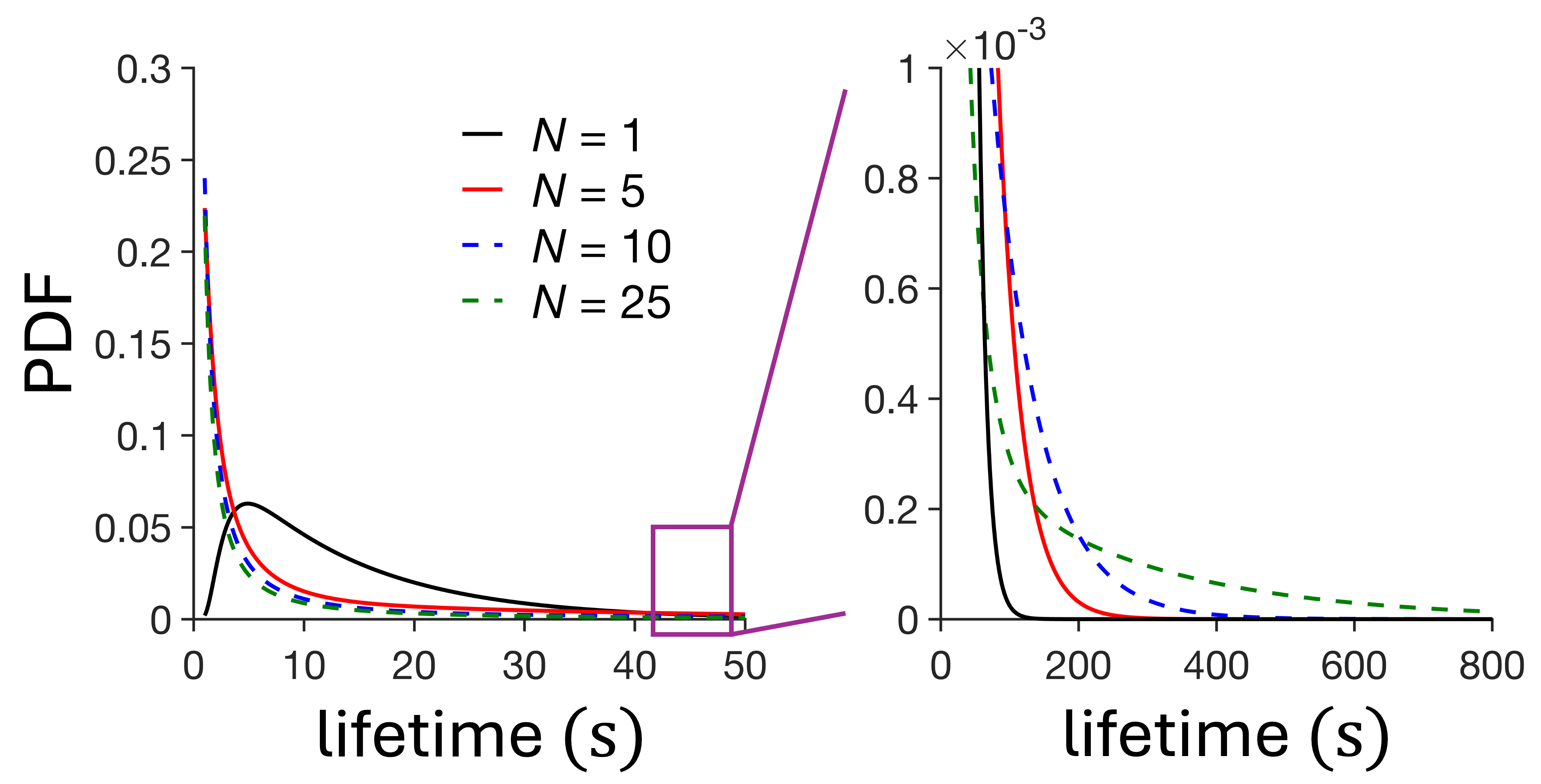}
\caption{The PDFs (probability distribution function) of lifetimes for different values of $N$. The mean lifetime $t_m$ is set to $\sim$15 s. The values of $\tilde{\mathcal{C}}$ are 0.034, 0.3, 0.6, and 1 for $N = $ 1, 5, 10, and 25, respectively. Here,  $\tilde{\mathcal{C}} = \mathcal{C}/\alpha$ and $\alpha$ represents a square unit of conformation changes per  second } 
\label{fig:lifetime_dstrs}
\end{figure}

For different values of $N$, the PDFs (probability distribution function) are shown in the first panel of Fig.~\ref{fig:lifetime_dstrs}, keeping the mean lifetime $t_m$ the same. Except for the exponential trend in $N=1$, the rest of the trends are power-law for $N>1$.  In the long time limit, the exponential trend in $N = 1$ decays much faster than $N>1$. As $N$ increases, the decay is slower (see the enlarged second panel in  Fig.~\ref{fig:lifetime_dstrs} ). The fat-tails in the long time limit for $N > 1$ complement the mean lifetime $t_m$  (the second panel in  Fig.~\ref{fig:lifetime_dstrs}), however, the probability of significantly large lifetimes is extremely low, irrespective of $N$. 

Using the form $S(t)$ in Eq.~\ref{eq:sur} for $N = 1$ may be inappropriate, as it implies only one conformation state and the continuum limit is invalid. Note that $N = 1$ is equivalent to the first-order kinetics, which must reflect an exponential trend. Surprisingly, the form $S(t)$ in Eq.~\ref{eq:sur} predicts an exponential trend for $N = 1$ and thus similar exponential trend by PDF $ = -\md S(t)/\md t$  (Fig.~\ref{fig:lifetime_dstrs}). Therefore, whether we use Eq.~\ref{eq:sur} for $N = 1$ or an exponential distribution function, the outcome remains the same.

\newpage
\subsection{Hybrid simulation method}

We use the Brownian dynamics (BD) to simulate the Brownian motion of individual molecules and their phase-separated clusters in a two-dimensional square box using periodic boundary conditions in both the $x-$ and $y-$ directions.  The coarse-grained protein molecules are treated as non-overlapping circular discs of radius $r_p = 5$ nm, and their motion is simulated using the Cichocki–Hinsen algorithm \citep{cichocki1990dynamic, smith2017macromolecular}. This chosen size of a molecule is not specific to any protein. Some miRNA proteins were previously reported to have a similar size range \citep{gu2012detection}. For simplicity, the electrostatic, other long-range, and hydrodynamic interactions among molecules and clusters are ignored. The simulation time step is set to a sufficiently small value $\tau_s = 10^{-5}$s so that overlap due to a large increment in position can be avoided in one time step. 

\begin{figure}
\centering
\includegraphics[width=0.5\linewidth]{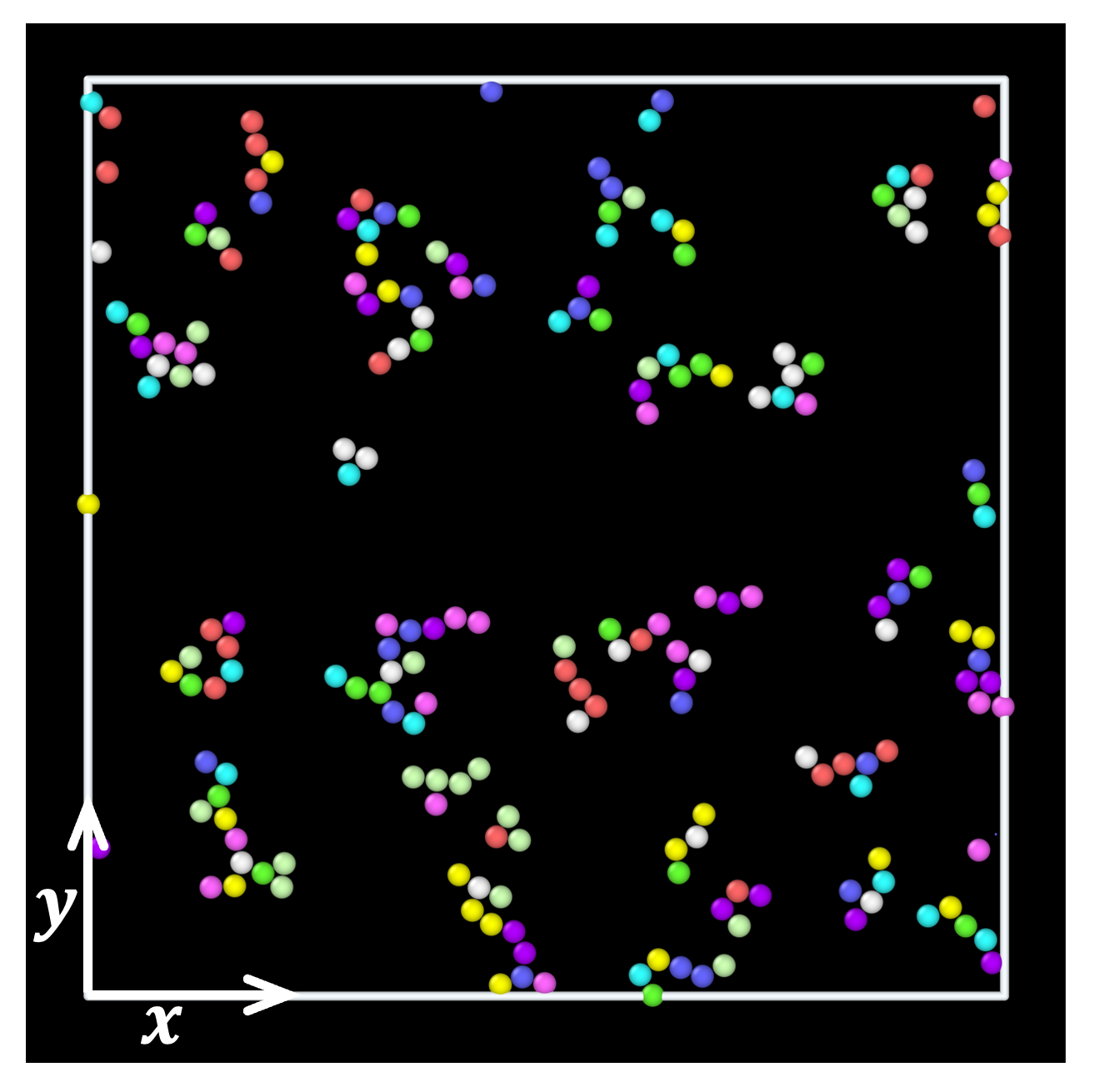}
\caption{A snapshot of Brownian dynamics simulation box containing coarse-grained molecules and their clusters.} 
\label{fig:bd_suppl}
\end{figure}

We simulate the motion of 200-400 molecules (Fig.~\ref{fig:bd_suppl}). The size of the simulation box is set so that a suitable area fraction (equivalent to a concentration) can be achieved.  It is well-known that concentration plays a role in phase separation. However, here, we do not aim to generate a phase diagram. Rather, our goal is specific to understanding the influence of interaction specificity on condensate features for a known concentration. That is why, for the time being, we choose two values of area fractions $a_f$, 0.05 and 0.1, arbitrarily to test our model.

The diffusivity of the individual protein molecules is set to $D_p = 0.1$ \textmu m$^2$/s. The value of diffusivity may appear very low. This low diffusivity has been reported earlier for RNA protein molecules inside the nucleus and cytoplasm \citep{tadakuma2006imaging,braga2007reaction,chen2014cytoplasmic}. However, it can be argued that the low diffusivity is because of the crowded and highly viscous environment inside the nucleus and cytoplasm.  The actual diffusivity would differ if not crowded. It is apparent that, in phase separation, diffusion plays a key role for transporting the protein molecules. In our simulations, though there is no crowding, to keep the time scale for diffusion of the same order mimicking the environment of the nucleus and cytoplasm, we choose such a low value of diffusivity.  

We use the sequential binding-unbinding model with the simulation framework mentioned above in a hybrid manner. When two protein molecules (whether part of clusters or individuals) are in close proximity, they bind with a sticking probability $p_s$ and remain in a bound state for a lifetime that is randomly drawn from the PDFs shown in Fig.~\ref{fig:lifetime_dstrs}. The PDFs are generated for ranges of parameters shown in table~\ref{tab:sim_table}. During the bound state, the molecules are part of clusters. The translational and rotational diffusivity of clusters are set to $D_c^t = D_p \, (r_p/r_c)$ and $D_c^r = D_c^t/(2 \, r_c^2)$, respectively, where $r_c$ is the radius of gyration of a cluster. 

\begin{table}[h!]
  \centering
  \caption{Simulation parameters}
  \label{tab:sim_table}
  \begin{tabular}{|c|c|c|c|}
    \hline
    mean time, $\tau_m$ & $N = 1$ & $N = 5$ & $N = 25$ \\
    \hline
    15s & $\tilde{\mathcal{C}} = 0.034$ & $\tilde{\mathcal{C}} = 0.3$ & $\tilde{\mathcal{C}} = 1$ \\
    \hline
    130s & $\tilde{\mathcal{C}} = 0.0038$ & $\tilde{\mathcal{C}} = 0.03$ & $\tilde{\mathcal{C}} = 0.047$\\
    \hline
  \end{tabular}
\end{table}

An accurate estimate of the initial sticking probability would be difficult, which would require a rigorous molecular dynamics simulation for a specific protein. However, here, we set $p_s = 0.1$ based on the following approximation. For a distance of $d$ between two discs, the Van der Waals force experienced by each would be $|F_{vdw}| \sim (A/12) \, (r_p/d^2)$. Therefore, the potential energy barrier  $\Delta U/(k_B T) \sim (F_{vdw} \, d)/(k_B \, T)$. Following Eyring's theory \citep{eyring1935activated,glasstone1941theory,ball1987universal}, the sticking rate would be 
\begin{equation}
k_s \sim (k_B T/h) \, {\rm exp} (-\Delta U/ (k_B T)).
\end{equation}
Here, $A$, $k_B$, $T$, and $h$ are the Hamaker constant, Boltzmann constant, temperature, and Planck's constant, respectively.  The sticking probability for two discs at a distance $d$ in one simulation time step is $p_s = 1 - {\rm exp} (\tau_s \, k_s)$. For $A = 10^{-19}$ J (protein in water), $r_p = 5$ nm, $d = 0.5$ nm, $T = 303$ K, the sticking probability yields to $p_s = 0.1$.  

Another variable in our simulations is the number of binding regions $n_b$ per protein disc in 2D simulations. For example, if $n_b = 3$, it implies that a protein disc can allow a maximum of 3 other protein discs to be its neighbors in the bound state. We perform simulations for two values of $n_b$, 3 and 4.  

\newpage
\begin{figure}
\centering
\includegraphics[width=1\linewidth]{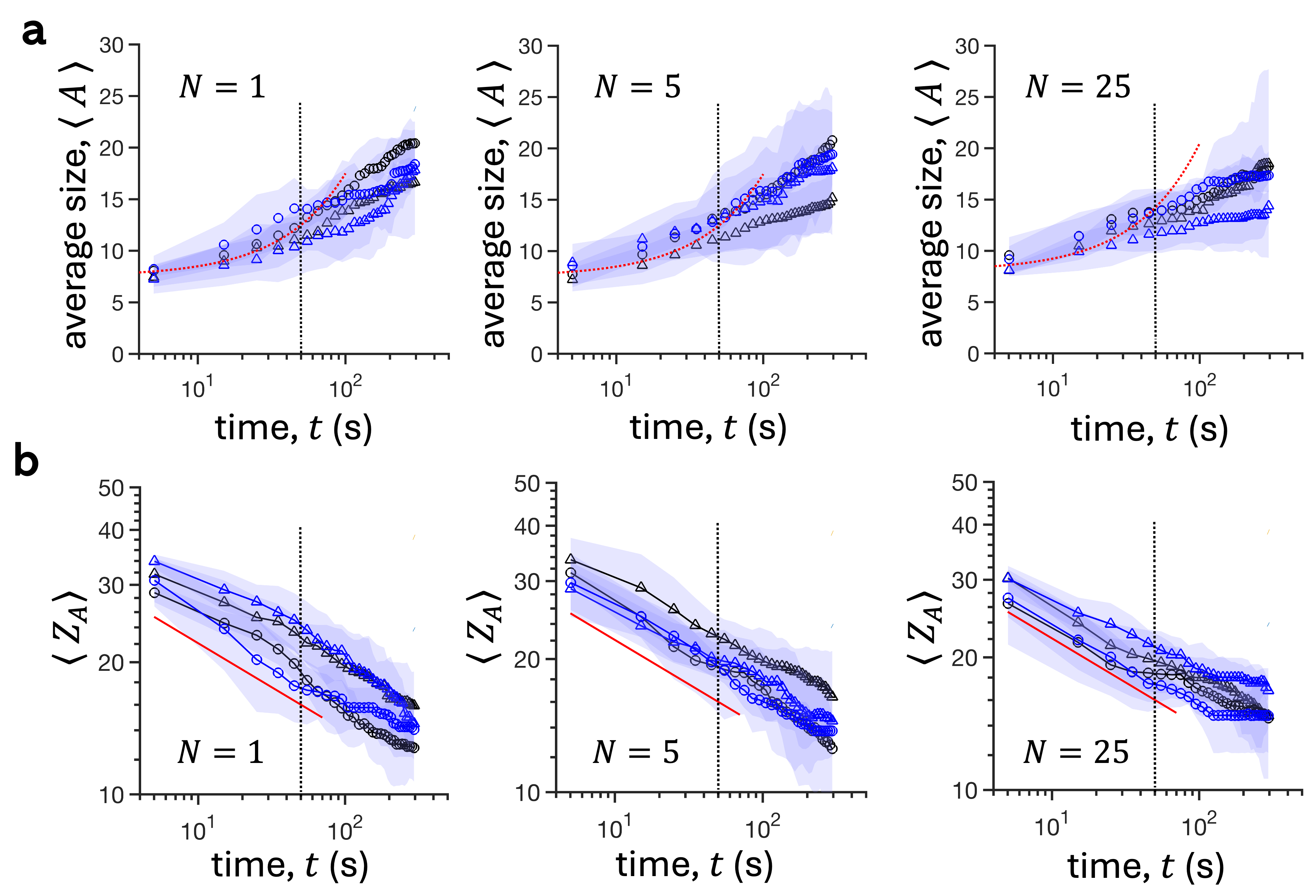}
\caption{Variation of the average size of clusters $\Aav$ (a) and the total number of clusters $\NA$ (b) with time $t$ for different values of the total number of probable conformational changes $N$. Here, the black and blue symbols are for the number of binding regions $\mathcal{R}_b = 3$ and $\mathcal{R}_b = 4$, respectively, and  $\circ$ and $\triangle$ represent the mean lifetime of the lifetime distributions $\tau_m = 15$ s and $\tau_m = 130$ s, respectively. Panels in (a) and (b) correspond to 
the area fraction $a_f = 0.05$. The red dotted curves in panel (a) fit the data points by $y = (mx + c)$ on the left of the black dotted vertical lines. The red solid lines in panel (b) represent the slope $-0.2$. The shaded regions represent the 95\% confidence limits from four independent simulations.} 
\label{fig:size_time_supp}
\end{figure}

\subsection{Counting number of neighbor exchanges}

\begin{figure}
\centering
\includegraphics[width=1\linewidth]{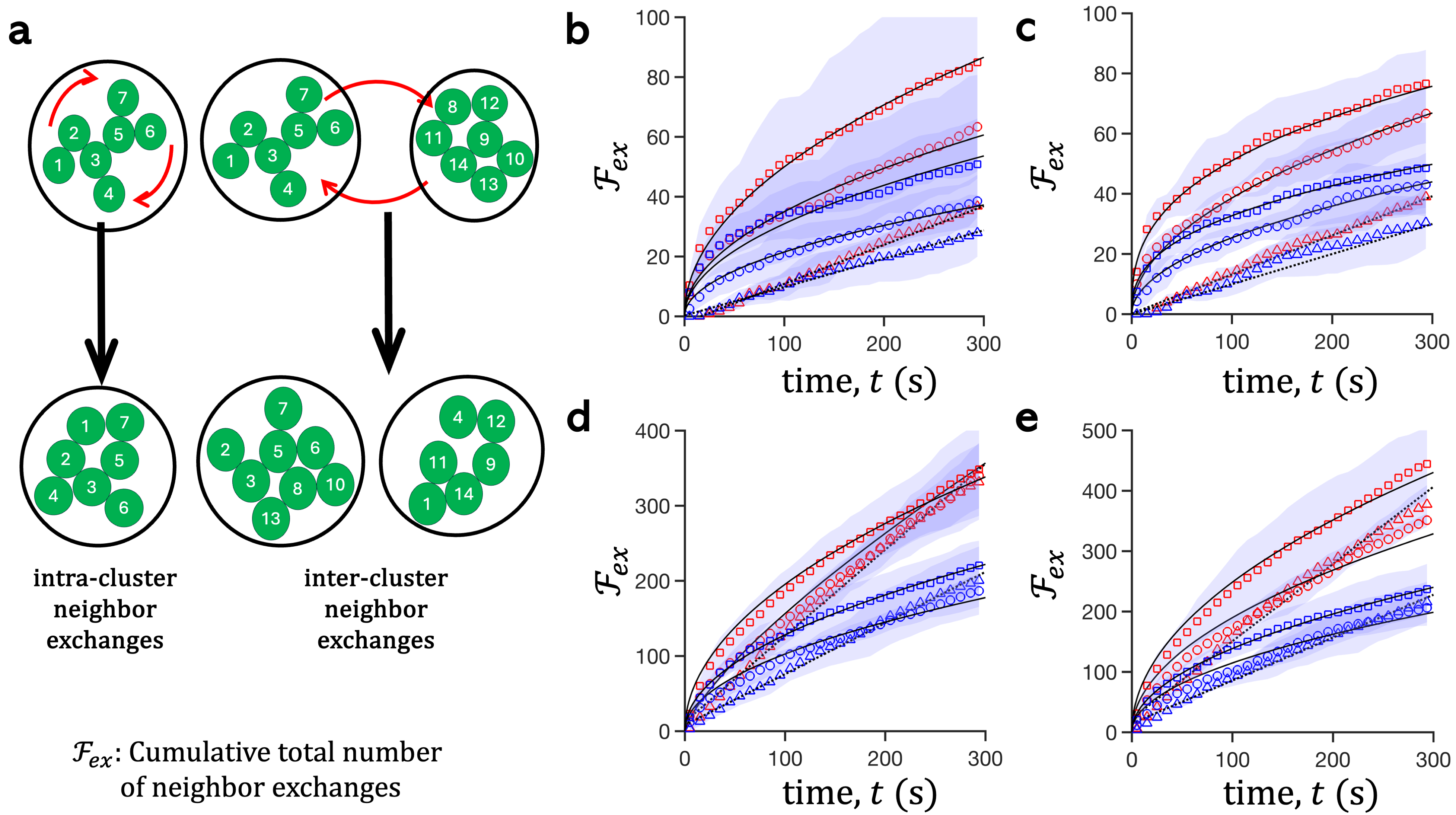}
\caption{(a) Schematic of different types of neighbor exchanges. Variation of the cumulative total number of exchanges $\mathcal{F}_{ex}$ with time $t$. Parameters: (b) area fraction $a
_f = 0.05$, the mean lifetime of distribution $\tau_m = 130$ s; (c) $a
_f = 0.1$, $\tau_m = 130$ s; (d) $a_f = 0.05$, $\tau_m = 15$ s; (e) $a_f = 0.1$, $\tau_m = 15$ s. In (b)--(e), red and blue symbols correspond to the number of binding regions $\mathcal{R}_b = 3$ and $\mathcal{R}_b = 4$, respectively. Here, $\triangle$, $\circ$, and $\square$ are for the total number of probable conformational changes $N = 1$, $N = 5$, and $N = 25$, respectively. The black solid curves are fit for $y = m x^{1/2}$ and black dotted lines for $y = m x$. The shaded regions represent the 95\% confidence limits from four independent simulations.} 
\label{fig:exchanges_time_suppl}
\end{figure}

Counting the number of intra- and inter-cluster neighbor exchanges is difficult and a tricky job. We apply a simple approach as follows. When a molecule $i$ is a part of a cluster, we track the ids $j$ of the molecules bonded with $i$. For each $i$, we compare two sets of $j$ collected at time $t$ and after a window time $(t + \Delta t)$. These sets must be non-empty (an empty set represents that $i$ is not part of a cluster). If these two non-empty sets are disjoint sets, which may imply that $i$ is a part of either its old or a new cluster, we count one neighbor exchange event corresponding to $i$. The former follows the definition of a pure neighbor exchange event at the scale of an individual molecule when it is a part of a cluster. In a similar manner, we perform this operation for rest of the molecules which are part of clusters, and count the total number of such events in that duration. We then obtain the cumulative total number of events at time (say, $(t+\Delta t)$), and this quantity is represented by $\mathcal{F}_{ex}$. The variations of $\mathcal{F}_{ex}$ with time for ranges of parameters, such as $N$ and $\mathcal{R}_b$, are shown in Fig.~\ref{fig:exchanges_time_suppl}.  

\bibliographystyle{vancouver}
\bibliography{References}